\documentclass[pre,aps,floatfix,notitlepage,10pt,twocolumn]{revtex4-1}
\usepackage{graphicx,epsfig}
\usepackage{bm}
\usepackage{dcolumn}
\usepackage{xcolor}% coloured text
\usepackage[breaklinks=true,colorlinks,citecolor=blue,linkcolor=blue,urlcolor=blue]{hyperref}

\usepackage[latin2]{inputenc}
\usepackage{ulem}
\usepackage{amsmath}
\usepackage{hyperref}

\newcommand{\doid}[1]{\href{https://doi.org/\detokenize{#1}}{doi: \detokenize{#1}}}
\newcommand{\ie}{\textit{i.e.}}
\newcommand{\eg}{\textit{e.g.}}
\newcommand{\etal}{\textit{et al.}}

\def\noi{\noindent}
\def\bc{\begin{center}}
\def\ec{\end{center}}
\begin{document}
\title{Cellular and Developmental Basis of Avian Structural Coloration}

\author{Vinodkumar Saranathan$^{1,2}$}\email{Vinodkumar.Saranathan@aya.yale.edu}
\author{Cédric Finet$^{1}$}
\affiliation{$^{1}$Division of Science, Yale-NUS College, 10 College Avenue West, 138609, Singapore.\\
$^{2}$NUS Nanotechnology and Nanoscience Initiative, National University of Singapore, 117576, Singapore.}

%\date{\today}

\begin{abstract}
\noi Vivid structural colors in birds are a conspicuous and vital part of 
their phenotype. They are produced by a rich diversity of integumentary 
photonic nanostructures in skin and feathers. Unlike pigmentary 
coloration, whose molecular genetic basis is being elucidated, little is 
known regarding the pathways underpinning organismal structural 
coloration. Here, we review available data on the development of avian 
structural colors. In particular, feather photonic nanostructures are 
understood to be intracellularly self-assembled by physicochemical 
forces typically seen in soft colloidal systems. We identify promising 
avenues for future research that can address current knowledge gaps, 
which is also highly relevant for the sustainable engineering of 
advanced bioinspired and biomimetic materials.\\

\noi \textbf{Keywords:} structural colors, biophotonic nanostructures, 
self-assembly, skin coloration, plumage coloration

\end{abstract}

\maketitle

\section*{Introduction}

\noi While pigmentary colors result from wavelength-selective molecular 
absorption and re-emission of light, vivid saturated colors are produced 
via physical or structural means, or sometimes a combination of both 
\cite{Vukusic2003,Cuthill2017}. Organismal structural colors arise from light scattering by 
biophotonic nanostructures with compositional variation (\ie, refractive index contrast) on the order of visible light wavelengths 
\cite{Vukusic2003,PCbook}. They can be further classified based on whether the 
scattering is \textit{incoherent} (\eg, white color), 
arising from uncorrelated or spatially independent scatterers (in 
Rayleigh, Tyndall or Mie regimes depending on particle size) or \textit{coherent}, as a result of constructive interference of light due to 
periodic or quasi-periodic spatial material variation with 
characteristic length scales of about 100-350 nm \cite{Vukusic2003,PCbook}. The latter 
class of interference colors, especially (ultra)violet, blue and green 
hues are quite conspicuous in animals and produced by a stunning 
diversity of underlying epidermal or integumentary photonic 
nanostructures \cite{Vukusic2003,Hill2006a,Saranathan2015}. They constitute a very important aspect of 
the appearance of animals including birds, as they are often used in 
aposematism, crypsis or in inter- and intra-sexual signaling \cite{Cuthill2017}\cite{Hill2006b}.

By contrast to pigment-based coloration and pattern formation, 
there is a dearth of developmental studies on organismal structural 
coloration in general, and their underlying genetic basis is only just 
unraveling \cite{Zhang2017,Airoldi2019}\cite{Morse2020}. This is in part, because most model species 
lack structural coloration or remain uninvestigated, even if present. In 
this review, we describe the progress to date in understanding the 
genetics and development of structural color production in birds, a 
cosmopolitan group of over 10,000 species with vibrant and diverse 
coloration \cite{Hill2006a,Hill2006b}. We identify promising avenues for future research 
that can address current knowledge gaps.

\section*{Structural coloration in avian skin}

\begin{figure*}
	\includegraphics[width=\textwidth,height=11.5cm]{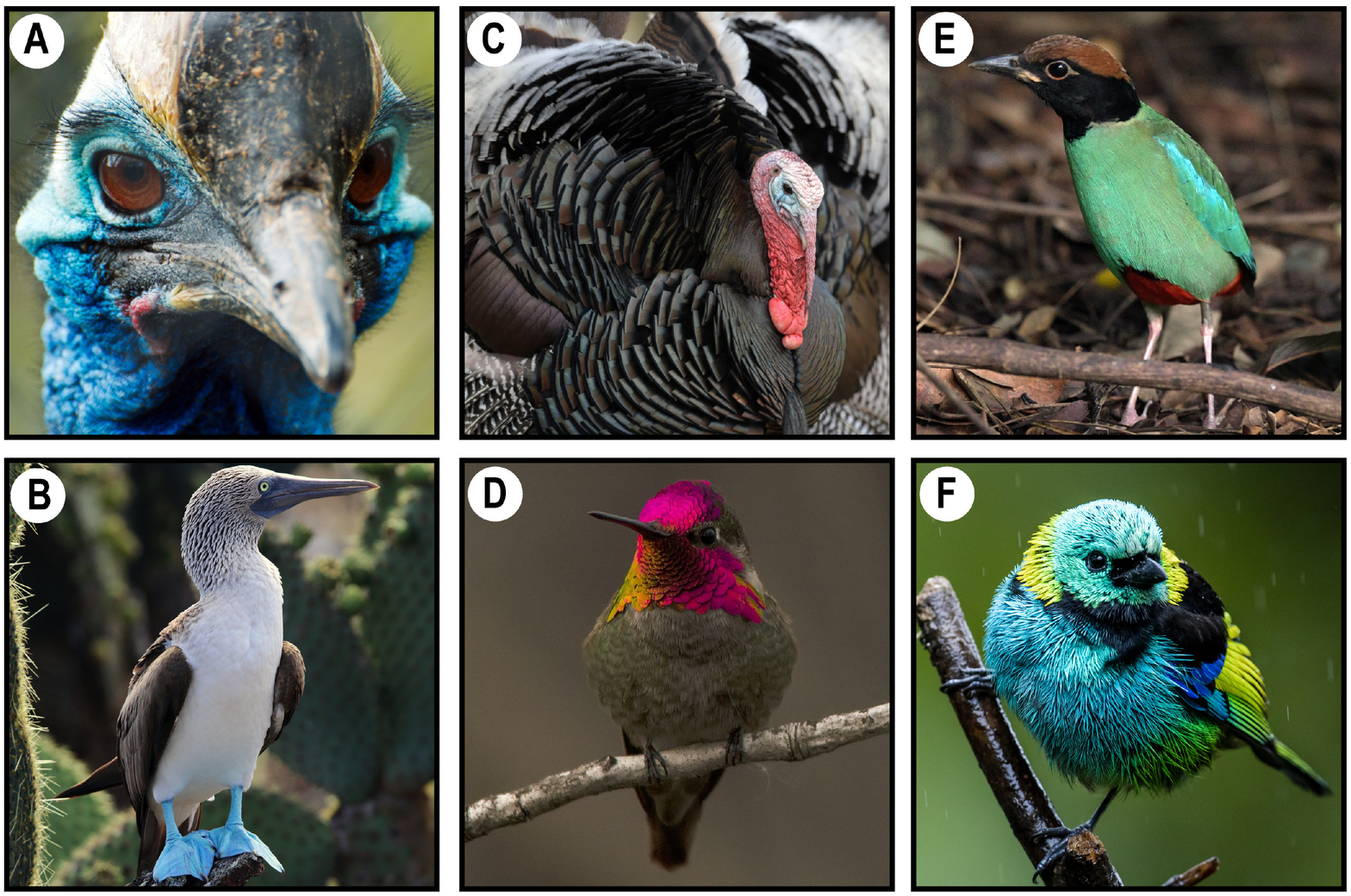}
	\caption{
		Diversity of Avian Structural Coloration. Non-iridescent Skin: (A) Southern Cassowary (\textit{Casuarius casuarius}), (B) Blue-footed Booby (\textit{Sula nebouxii}); Iridescent Feather Barbule: (C) Wild Turkey (\textit{Meleagris gallopavo}), (D) Anna's Hummingbird (\textit{Calypte anna}); Non-iridescent Feather Barb: (E) Hooded Pitta (\textit{Pitta sordida}), (F) Green-headed Tanager (\textit{Tangara seledon}).\\ 
		Photo Credits: A - Lucy Bridges, B - Rene Leubert, D - Daniel Arndt, F - 
		Nick Athanas, C and E - Vinodkumar Saranathan (All images are cropped and reproduced under CC-BY-NC-SA 2.0)
		\label{fig1}}
\end{figure*}

Non-iridescent structural colors prominently occur in bare skin 
(especially around the eye), bill (ramphotheca) and feet (podotheca) and 
has convergently evolved in over 50 bird families, likely driven by 
sexual selection \cite{Prum2003,Prum2006} (Figs. \ref{fig1}A-B). They are produced by 
constructive light interference from 2D quasi-periodic arrays of 
parallel collagen (Refractive Index 1.42) fibrils in the 
mucopolysaccharide matrix (RI 1.35) of the dermis and underlain by a 
layer of melanin granules called melanosomes \cite{Prum2003,Prum2006} (Fig. \ref{fig2}A). 
The collagen fibrils are in turn bundled into larger macrofibrils 
(fibers), tens of microns in diameter and several hundred microns in 
length that are apparently produced by a single collagenocyte in such a 
way that their longitudinal axis is aligned nearly parallel to the skin 
surface \cite{Prum2003}. However, the Velvet Asity (\textit{Philepitta 
castanea}) uniquely among birds has evolved a 2D photonic crystal (PC) 
\cite{PCbook} analog with ordered hexagonal arrangement of collagen fibrils 
that is derived from ancestral quasi-periodic state found in the sister 
sunbird asities (\textit{Neodrepanis spp.}) \cite{Prum2003,Prum1999}. 
This transition is apparently driven by female preference for highly 
saturated hues in this lek polygynous species rather than for 
directional optical properties, as the overall papillose geometry of the 
facial caruncle and the polycrystalline nature of macrofibrils attenuate 
iridescence \cite{Prum2003,Prum1999}.

The genetic and developmental basis of structural color 
producing dermal collagen arrays remains unstudied. Nevertheless, Prum 
and coworkers \cite{Prum2003} have suggested some prerequisites for the 
evolution of photonic dermal collagen arrays. In a plausible sequence, 
these include -- loss of feathers exposing bare skin (apteria), 
thickening of dermis with a concomitant proliferation of collagen arrays 
to increase scattering efficiency given their low refractive index 
contrast, dermal melanization to absorb any unscattered light that would 
otherwise wash out structural hues, and near-uniform specification of 
larger than usual fibrils ($>$ 100 nm) to make avian visible hues. It is 
conceivable that evolution of bare skin in birds is accompanied by 
dermal thickening for mechanical reasons and melanization as protection 
against UV damage. 

Here, we focus on the molecular basis of collagen fibrillogenesis, which 
ultimately determines the photonic aspect of dermal collagen arrays 
(Fig. \ref{fig2}A). The hierarchical assembly of triple-helical collagen proteins 
that occurs in the dermal extracellular matrix (ECM) into collagen 
fibres is intrinsic and integral to the structure and function of 
vertebrate connective tissues, including dermis \cite{Kadler2008}. Collagens, 
specified by the diverse COL gene family, are ubiquitous and largely 
conserved across vertebrates, although birds have a slightly reduced 
diversity \cite{Haq2019}. The fact that collagen synthesis and fibrillogenesis 
in vertebrates are tightly regulated \cite{Kadler2008}, with large irregularly 
sized fibrils seen pathologically \cite{Liu1997}, suggests that photonic 
organization of fibrillar bundles likely stems from molecular regulation 
of fibril diameter, inter-fibril spacing and arrangement. One promising 
regulator is Tenascin-X (TNX; Tenascin-Y in birds \cite{Hagios1996}) with 
epidermal growth factor (EGF) and fibronectin (FN) domains, which is 
implicated in determining inter-fibril spacing as well as accelerating 
rate of fibril formation by interacting with ECM proteoglycans such as 
decorin \cite{Kadler2008}. Interestingly, loss of either decorin \cite{Zhang2009} or TNX 
\cite{Minamitani2004} results in irregularly arranged, large-diameter fibrils. The 
ETS family transcription factor FLI1 (with known avian homolog) is also 
of interest as it represses fibrillar collagen genes, while upregulating 
the production of small leucine-rich proteoglycan (including decorin), 
during fibrillogenesis \cite{Asano2009}. However, spatio-temporal changes in 
fibrillar and fibril-associated collagen expression could result in 
similar changes in phenotype \cite{Liu1997}, suggesting that family-specific 
differences in collagen fibril composition could also be responsible for 
the repeated evolution of this trait in birds \cite{Haq2019}.

\section*{Structural coloration in avian plumages}

\subsection*{Iridescent Feather Barbule Coloration}

\begin{figure*}
	\includegraphics[width=\textwidth,height=11.5cm]{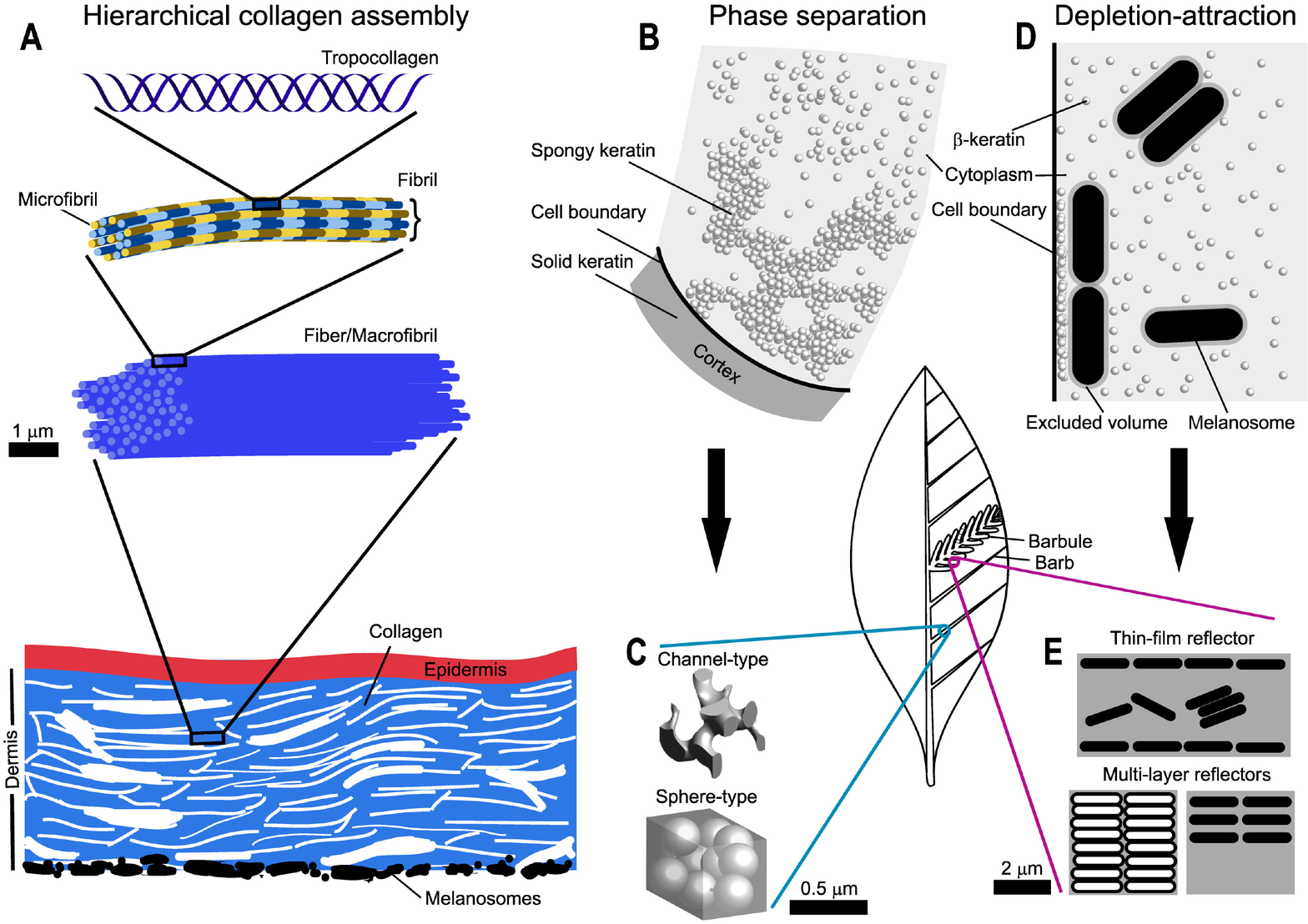}
	\caption{A Schematic for the Assembly of Photonic 
	Nanostructures in Bird Skin (A) and Feathers (B-E). (A) In hierarchical 
	collagen fibrillogenesis \cite{Kadler2008} depicted here in increasing levels of 
	spatial organization, diverse triply-helical collagen molecules assemble 
	into composite microfibrils that organize into macrofibrils/fibers, 
	which are synthesized by a single dermal corneocyte \cite{Prum2003}. The lower 
	panel shows a dense packing of dermal collagen fibres (blue) oriented 
	parallel to the skin surface with occasional gaps (white), and underlain 
	by a layer of melanosomes. The fibril size, spacing and organization are 
	molecularly determined \textit{in vivo}, even though collagen can 
	spontaneously self-assemble into fibrils \textit{in vitro} \cite{Kadler2008}. 
	(B) Phase separation of $\beta$-keratin is hypothesized to result in 
	spontaneous stereotypical pattern formation (\eg, dendritic spinodal-like patterns depicted) at the periphery of barb cells late in development, when there is a "super-critical" 
	concentration of $\beta$-keratin near cell edges due to capillary flow 
	\cite{Prum2009}. (C) Two types of photonic glasses are known in feather barbs 
	- channel- and sphere-type, which are analogous to morphologies seen 
	during spinodal decomposition or nucleation-and-growth \cite{Saranathan2012}. (D) 
	During iridescent feather barbule development, melanosomes are thought 
	to aggregate, driven by depletion-attraction forces \cite{Maia2012,Shawkey2015}. Note, 
	a small amount keratin molecules trapped in-between the cell membrane 
	and assembling melanosomes could form the thin cortex of iridescent 
	barbules. Timing of keratin synthesis and polymerization within barbule 
	keratinocytes is key to enable the self-assembly of melanosomes 
	\cite{Durrer1967}. (E) Melanosomes can be solid or hollow, and can form a 
	thin-film monolayer, densely close-packed or periodically-spaced 
	multilayers. Approximate scale bars are shown alongside biophotonic 
	nanostructures to illustrate the typical spatial periodicities (150-300 nm) needed for light interference.
	\label{fig2}}
\end{figure*}

Structural coloration in feather barbules is generally iridescent (Figs. \ref{fig1}C-D) and produced by interference from biological analogs of 1DPCs 
\cite{PCbook} -- thin-film or multi-layer (spaced or close-packed) arrays of 
melanosomes (RI $\sim$ 2.0) embedded in a $\beta$-keratin medium (RI 1.58) 
\cite{Prum2006} (Fig. \ref{fig2}E). However, considerable systematic variation exists 
in the morphology and arrangement of melanosomes within iridescent 
barbules \citep[See Fig. 3 of ][]{Durrer1986}\cite{Gruson2019,Maia2013}. The melanosomes can be spherical, 
lozenge to rod-shaped about 1-2 $\mu$m in length with various aspect ratios, 
or pancake-shaped. The melanosomes involved in barbule iridescence are 
usually comprised of eumelanin, although some pigeons (\textit{Columba trocaz}) utilize phaeomelanins \cite{Prum2006}. Some taxa 
(starlings, hummingbirds and quetzals) have evolved arrays of novel 
hollow/air-filled (RI 1.0) rod- and pancake-shaped melanosomes from 
ancestral solid types, leading to an increased refractive index contrast 
and thereby, extravagant interference colors \cite{Durrer1986}. Some have evolved 
close-packed 2DPC-like \cite{PCbook} square (peafowl) or hexagonal (ducks and trogons) arrangement of melanosomes \cite{Prum2006}\cite{Durrer1986}, although optically 
they seem to function as multilayers \cite{Stavenga2017,Freyer2019}. These diverse 
melanosomal arrays have convergently evolved numerous times across birds
and show complex evolutionary history \cite{Prum2006}, with both 
solid and hollow melanosomes in different plumage patches in some 
species \cite{Gruson2019}. 

Published more than 50 years ago, Durrer and Villiger's 
\cite{Durrer1967} description of hollow melanosome assembly in barbules of a 
starling (\textit{Lamprotornis}) still remains an authoritative 
source on the ontogeny of iridescent barbule coloration. During feather 
development, melanosomes produced within specialized melanocytes are 
dendritically transferred to the barbule plate keratinocytes via 
endocytosis-like process \cite{Durrer1986,Durrer1967}. Melanosomes in developing 
non-iridescent barbules have an exclusion zone around them and seem 
randomly embedded in a matrix of rapidly polymerizing keratin that 
physically prevents the migration of melanosomes to the cell boundary 
\cite{Durrer1967}. Whereas, in iridescent barbules, $\beta$-keratin is proliferating 
and polymerizing into small fibrils that do not fuse and remain confined 
to the center of the cell, with the cytoplasmic melanosomes free to 
diffuse to the cell membrane. The melanosomes are eventually 
mechanically confined by the keratin mass as it grows to fill the cell 
volume, and ordered into a marginal monolayer as the cell flattens upon 
death and dehydration \cite{Durrer1967}. A relatively recent developmental 
study in Blue-black Grassquit (\textit{Volatinia jacarina}) observed 
a greater density of larger, more uniformly-sized melanosomes in 
barbules of iridescent males relative to non-iridescent females 
\cite{Maia2012}. Based on their observations that the organization of the 
melanosomes into a flat monolayer occurs late in barbule development, 
when the cell is dying, they proposed that depletion-attraction forces 
re-organize melanosomes into a monolayer, as opposed to cellular or 
molecular mechanisms \cite{Maia2012,Ghosh2015}. When melanosomes aggregate, the volume 
that keratins are normally excluded from occupying is reduced, 
increasing entropy and lowering the free-energy of the system (Fig. \ref{fig2}D).

In \textit{Lamprotornis sp.} \cite{Durrer1967}, hollow melanosomes 
are already formed within melanocytes before being transferred to 
keratinocytes. Premelanosomes, large vesicles ($\sim 1 \mu$m diameter) of Golgi 
origin filled with fine granular tyrosinase are rapidly generated within 
melanocytes, and later incorporate many small vesicles from the 
cytoplasm. A zigzag 5-6 layer lamellae forms centrally around which the 
smaller vesicles are organized like beads on a string. As the 
premelanosomes flatten, melanin is rapidly synthesized around the 
foam-like central matrix, which eventually becomes the air-filled 
internal structure of hollow melanosomes. By contrast, in regular solid 
melanosomes, melanin synthesis occurs at the zigzag lamellae and 
completely fills the premelanosome \cite{Birbeck1962}. Another recent study on the 
ontogeny of hollow melanosomes in iridescent barbules of Wild Turkey (\textit{Meleagris gallopavo}) indicated that melanosomes are mostly 
solid and randomly oriented during transport \cite{Shawkey2015}. Once inside the 
barbule keratinocyte, however, they are mostly oriented in the same 
direction, and electron-dense material from the core is lost, prior to 
being close-packed into a hexagonal lattice. Shawkey \etal 
\cite{Shawkey2015} suggest that in turkeys, melanosomes could be a composite with 
a phaeomelanin core and eumelanin mantle, and that the chemically 
unstable phaeomelanin core can degrade upon changes in local environment 
(\eg, pH). That the mechanisms for hollow melanosome 
ontogeny are convergent is not surprising given iridescent barbule 
coloration using hollow melanosomes has evolved numerous independent 
times across birds \cite{Prum2006}\cite{Gruson2019,Maia2013}. 

Several outstanding questions remain. Although it is becoming 
clear that melanosomes late in development assemble via a "\,crowding 
mechanism" \cite{Ghosh2015}, this needs to be reconciled with spatiotemporal 
changes in keratin synthesis and polymerization in barbule keratinocytes 
\cite{Durrer1967}, which can affect the position of the melanosomes relative to 
the cell boundary. Furthermore, depletion-attraction alone cannot 
explain how spaced multi-layers form (Fig. \ref{fig2}E), let alone more complex 
arrays with double-layered melanosomes in Birds of Paradise 
\cite{Prum2006}\cite{Durrer1986}. Some authors have argued that the high aspect ratios of 
melanosomes in iridescent barbules relative to non-iridescent ones is 
enough for the self-emergence of layering \cite{Shawkey2015,Norden2019}. However, the 
observed melanosome aspect ratios seem far from the optimum under 
granular packing considerations of spherocylinders \cite{Baule2013}. Moreover, 
melanosome morphology can vary in a single species more than previously 
appreciated \cite{Prum2006}\cite{Durrer1986,Gruson2019}, further confounding these analyses.

Given the key role of melanosomes in iridescent color 
generation, it is conceivable that convergent regulatory changes in 
melanin synthesis pathway have led to the repeated evolution of barbule 
iridescence in birds. The genetics of melanin-based coloration is 
well-studied in animals \cite{Orteu2020}, and recent progress in butterfly 
coloration suggest changes to single master regulatory genes can 
pleiotropically induce structural coloration \cite{Zhang2017}, while loss of 
melanin pathway genes can alter the gross morphology of scales 
themselves \cite{Matsuoka2018}. We wonder if similar changes could pleiotropically 
affect keratin expression and feather morphology in birds. The role of 
Melanocortin-1 Receptor (MC1R) in determining melanin patterning is 
inconsistent, but non-coding and coding differences in its repressor, 
agouti signaling protein (ASIP), is functionally significant across 
organisms \cite{Orteu2020}\cite{Funk2019}. A recent comparative genomic study on birds of 
paradise with extravagant iridescent barbule coloration \cite{Prost2019} 
suggests other promising candidates under putative positive selection -- 
ADAMTS20, implicated in melanocyte development through KIT ligand 
functioning, and ATP7B, involved in copper transport, an element 
essential for melanogenesis. A similar study in galliforms \cite{Gao2018} 
recovered functional changes in four melanogenesis genes, two of which 
are \textit{KIT} and \textit{ASIP}, but these were not specific to 
iridescence. More interestingly, they found difference in $\beta$-keratin gene 
expression between white and iridescent green feathers. Recently, 
melanocytes themselves have been found to autonomously determine color 
patterning and ASIP expression in adjacent dermal tissue, and this could 
be investigated via melanocyte transplantation from iridescent to 
non-iridescent barbule plates \cite{Funk2019}.

\subsection*{Non-iridescent Feather Barb Coloration}

By contrast to iridescent barbule colors (Figs. \ref{fig1}C-D), structural 
colors in feather barbs are usually non-iridescent (Figs. \ref{fig1}E-F) and 
have evolved in over 45 families across 12 bird orders \cite{Prum2006,Saranathan2012}. 
Two main classes of 3D glassy or quasi-periodic photonic nanostructures 
are recognized in spongy barb medullary cells (Fig. \ref{fig2}C) -- interconnected 
networks with anastomosing air channels and $\beta$-keratin rods 
(channel-type), and random close-packed arrays of spherical air voids in 
a $\beta$-keratin matrix (sphere-type). Some species with slate to blue-gray 
plumages have evolved rudimentary barb structural coloration produced by 
highly disordered versions of channel- and sphere-type nanostructures 
\cite{Saranathan2012}. However, in Little Penguin (\textit{Eudyptula minor}), 
non-iridescent blue barb colors are uniquely produced by 2D 
quasi-periodic bundles of parallel $\beta$-keratin fibres in air \cite{Dalba2011}. 
Recently, Saranathan \etal \cite{Saranathan2020} discovered at least 3 
parallel transitions in barb nanostructures of leafbirds, tanagers, and 
manakins from ancestral 3D photonic glasses to derived 3DPCs, apparently 
driven by female preference for highly saturated colors \cite{Prum1999}, rather 
than directional signaling, as the random Pointillist presentation of 
the crystal domains reduces iridescence.

The channel- and sphere-type barb nanostructures (Fig. \ref{fig2}C) are 
remarkably similar to highly stereotypical morphologies seen during 
phase separation of binary mixtures via spinodal decomposition and 
nucleation-and-growth, respectively \cite{Saranathan2012}\cite{Dufresne2009}. Beyond structural 
analogies, there is growing evidence that barb photonic nanostructures 
are self-assembled within spongy medullary cells by visco-elastic phase 
separation of $\beta$-keratin from cytoplasm (Fig. \ref{fig2}B), followed by a dynamic 
self-arrest as a result of the competition between coarsening and 
cross-polymerization of neighboring keratin fibrils \cite{Saranathan2012}\cite{Saranathan2020,Dufresne2009}. The 
lone ground-breaking study on the development of spongy barb 
nanostructures along maturation gradients in a growing feather germ of 
Blue-and-Yellow Macaw (\textit{Ara Ararauna}) is consistent with the 
phase separation hypothesis \cite{Prum2009}. Around mid-development, 
medullary barb cells develop small electron-lucent, fluid-filled 
droplets in the center that coalesce and grow in a manner reminiscent of 
capillary transport seen in drying coffee-ring stains \cite{Prum2009}. At 
this stage, electron-dense granules (likely keratohyalin) are seen near 
the edges of barb keratinocytes \cite{Prum2009}\cite{Matulionis1970}. Once the droplets grow to 
fill most of the cell volume pushing all electron-dense materials (\eg, nucleus) to the periphery, $\beta$-keratin synthesis and 
polymerization commences from cell edges. The first $\beta$-keratin structures 
to form is a reticulate matrix of solid keratin fibres with a distinct 
"\,crown-of-thorns" appearance, very similar to those seen at the 
periphery of hollow medullary cells in white feathers, but they are too 
large to make a visible interference color. Interestingly, the 
development of non-iridescent feather barbs seems identical up until 
this stage, when cells normally die leaving behind the characteristic 
foam-like, pneumatic medulla \cite{Matulionis1970}. In photonic barbs, however, 
development proceeds further and in just over a few hours, the 
characteristic channel-type network spontaneously appears at the cell 
boundaries from a cytoplasmic background filled with granular material 
whose sizes corresponds well with RNPs \cite{Matulionis1970,Shin2017}. As this 
polymerizing spongy network grows, the volume occupied by the 
electron-lucent droplet shrinks. During this process, the barb cells 
have not yet apoptosed since nuceli remain visible \cite{Prum2009}. When the 
cells finally die, the cytoplasm, nuclei and other cellular machinery 
are replaced by air. During this entire process, neither membrane nor 
cytoskeletal templates or prepatterns were observed directing the 
assembly of $\beta$-keratin into spongy networks, consistent with a phase 
separation process \cite{Prum2009}. Turing-type patterning, which produces 
(quasi-)periodic stripes and spots, is another unlikely alternative, as 
this process usually occurs in 2D (not 3D), and often breaks down with 
growth over time \cite{Saranathan2020,Maini2012}. Whereas, liquid-liquid phase separation 
within cells is a growing paradigm to explain the fundamental 
organization and functioning of cells, including how RNPs can lead to 
the development of fibrous, self-organized pathologies \cite{Shin2017}. It is 
plausible that birds have co-opted such innate cellular processes for 
photonic self-assembly, and future studies will have to investigate the 
identity and function of RNPs during barb development \cite{Prum2009}\cite{Matulionis1970}. 

Feather development and genetic basis of keratin expression are 
typically studied in Chicken (\textit{Gallus gallus}), Japanese 
Quail (\textit{Coturnix japonica}) and Zebra Finch (\textit{Taenopygia guttata}) none of which have barb structural coloration, while homology relationships between and among $\alpha$- and $\beta$-keratins are 
only just being uncovered \cite{Greenwold2014,Ng2018}. Across birds, there is extreme 
variation in copy numbers of $\beta$-keratin genes (6 in owls - 149 in Zebra 
Finch; average 34). A complex pattern of differential expression of 
different types of keratin genes (scale and claw, feather, keratinocyte) 
from multiple chromosomal loci in different feather tissues has been 
documented \cite{Greenwold2014,Ng2018}, with many of the feather $\beta$-keratin genes 
evolving their own chromosome-wise transcription factors \cite{Bhattacharjee2016}. 
Nevertheless, barb ridge specific (\eg, barb \textit{vs.} barbule) expression profiles are unavailable \cite{Ng2018}. We believe the key to unlocking the molecular basis for $\beta$-keratin self-assembly lies in 
generating tissue-specific, time-resolved expression of feather and 
feather-associated keratins, and identifying copy number variation 
unique to each family/genus \cite{Gao2018}\cite{Ng2018}\cite{Bruders2020}. Different family-specific 
combinations of $\beta$-keratins might confer different macromolecular 
properties (\eg, hydropathy, charge) that can aid or hinder self-assembly, while tuning the stoichiometry of expressed keratins may predictably determine the length scale at which the phase 
separation arrests, which is crucial for color production. Comparatively 
studying the tissue-specific molecular structure of feather keratins or 
looking for differences in the macro-molecular packing of $\beta$-keratin 
filaments, for instance, in spongy photonic barbs \textit{vs.} hollow 
white barbs or barbules could also help illuminate the 
structure-function relationships underpinning keratin self-assembly.

\section*{Conclusions and Future Directions}

Major aspects of organismal structural color pathways remain currently 
opaque. Nevertheless, we have highlighted how birds appear to have 
co-opted developmental programs behind collagen fibrillogenesis in 
dermis, melanosome synthesis and inclusion in feather barbules, and 
keratin polymerization in feathers barbs, to produce structural 
coloration. The redundant regulatory control of fibrillogenesis, 
melanosynthesis and cornification provides alternative pathways that can 
be modified by selection. This could explain the repeated convergence of 
avian structural coloration. We have also discussed the important role 
played by short-ranged attractive and long-ranged repulsive forces 
typically seen in soft colloidal systems \cite{Morse2020}\cite{Ghosh2015}\cite{Shin2017}\cite{Ji2015}, in feather nanostructure development. 

Interrogating molecular regulation of collagen fibril spacing by 
manipulating collagen fibrillogenesis \cite{Kadler2008} is one future challenge 
that is tractable in Silky Chicken (artificially-selected variant of 
model \textit{Gallus gallus}), with its hypermelanized dermis and 
unique blue earlobes. Another direction is to determine the exact role 
played by high melanosome densities in self-assembly during iridescent 
barbule development \cite{Maia2012}. The putative function(s) of cytoplasmic 
RNPs in flocculation and/or self-arrest also needs to be investigated 
\cite{Shin2017}\cite{Ji2015} Comparative transcriptomics and genome-wide association 
studies might represent promising complementary approaches to cell and 
developmental biology. Two of $\sim$50 bird species with published complete 
genomes have barb structural coloration, while at least a quarter have 
barbule iridescence \cite{Zhang2014}. As successfully demonstrated for 
pigmentary coloration \cite{Gao2018}, these methods could help identify 
candidate genes, whose functions can be tested using latest 
genome-editing technologies (\eg, CRISPR-Cas9) in existing 
model species with iridescence -- Silky, Domestic Chicken, and Turkey.

A burgeoning number of studies in physics and engineering are looking to 
organismal structural coloration, which have been evolutionarily 
optimized over millions of years of selection, as a rich reservoir for 
the bioinspired design and synthesis of functional materials, given 
current challenges in sustainable manufacture and synthetic 
self-assembly at visible optical lengthscales \cite{Morse2020}\cite{McDougal2019}. Genetic and 
developmental knowledge of biophotonic nanostructures may lead to 
next-generation technologies that directly biomimic \textit{in vivo}
self-assembly using bio-similar and biodegradable materials \textit{in vitro}.

\begin{acknowledgments}

VS acknowledges support from Yale-NUS start-up funds (R-607-261-182-121) 
and a NRF CRP Award (CRP20-2017-0004), and is grateful to Eric Dufresne, 
Antónia Monteiro, Dan Morse, and Rick Prum for stimulating discussions 
over the years.\\

\noi\textbf{Declaration of Interests:} The authors declare no conflict of interests.

\end{acknowledgments}

\renewcommand{\bibpreamble}{
\textbf{References and Recommended Reading}\\
\\Papers of particular interest, published within the period of review, have been highlighted as:\\
* of special interest\\
** of outstanding interest\\	
}

%code to drop the [] around reference number labels%
\makeatletter 
\renewcommand\@biblabel[1]{#1} 
\makeatother

\setlength{\labelsep}{1.2em}

\clearpage

\section*{Box 1: Glossary}

\textbf{Refractive Index} -- this dimensionless metric describes the 
amount of retardation of light in a dense medium relative to vacuum. The 
RI determines the extent to which light rays are bent (refraction) when 
entering a dense material from air. See \cite{PCbook}.\\

\textbf{Biophotonic Nanostructure} -- a nanoscale feature in the animal 
integument with random, quasi-periodic or periodic variation in material 
composition that leads to a refractive index contrast (\eg
, $\beta$-keratin and air). See \cite{Vukusic2003}\cite{Prum2006}.\\

\textbf{Photonic Crystal} -- A concept borrowed from solid-state 
physics, a photonic crystal describes periodic variations in material 
composition that interferes with the propagation of light through the 
material. See \cite{Vukusic2003,PCbook}.\\

\textbf{1D, 2D, 3D} (photonic materials) -- The designation 1D, 2D or 
3D describes whether the material varies in composition or refractive 
index along one (\eg, a thin-film or multilayer lamellae), 
two (\eg, square or honeycomb lattices of cylindrical 
holes), or three (\eg, gem opals and cubic crystals) 
principal orthogonal directions (\eg, Cartesian axes). For 
instance, in a 2D columnar or fibrillar nanostructure, there is no 
variation along column/fibril axis (say \textit{z}-axis), but there 
is in the cross-sectional plane (along \textit{x} and \textit{y}). See \cite{Vukusic2003,PCbook}\cite{Prum2006}.\\

\textbf{Iridescence} -- a change in hue with angle of illumination or 
angle of observation.\\

\textbf{Self-assembly} -- a ubiquitous phenomenon that describes the 
spontaneous or emergent organization of materials at macro-molecular 
length scales, usually driven by thermal fluctuations and interactions 
among molecules.\\

\textbf{Phase separation} -- a process that describes the spontaneous 
unmixing of immiscible binary or ternary mixtures, under unfavorable 
conditions. See \cite{Shin2017}.\\

\textbf{Depletion-attraction} -- a process that describes the 
attractive force that brings together larger colloidal particles that 
reduces the excluded volume around the large particles that a competing 
smaller solute (depletant) cannot normally occupy, thereby increasing 
the positional entropy for the depletants and lowering the overall 
free-energy of the system. Depletion-attraction can also lead to phase 
separation. See \cite{Ji2015}.\\

\end{document}